\documentclass[authoryear]{elsarticle}

\def\astrobjca{$\alpha$ Cen A}
\def\astrobjcb{$\alpha$ Cen B}
\def\astrobjp{Procyon A}
\def\astrobjb{$\eta$ Bootis}
\def\astrobjh{$\mu$ Her}
\usepackage{natbib}

 \usepackage{graphicx}
\usepackage{amssymb}

\journal{New Astronomy}
\begin{document}

\begin{frontmatter}



\title{Models of $\mu$ Her with asteroseismic constraints}


\author[auth1]{Wuming Yang}
\ead{wuming.yang@hotmail.com}
\author[auth2]{Xiangcun Meng}
\address[]{School of Physics and Chemistry, Henan Polytechnic
              University, Jiaozuo 454000, Henan, China. }

\begin{abstract}
 Using the Yale stellar evolution code, models of \astrobjh{} based on
 asteroseismic measurements are constructed. A $\chi^{2}$ minimization
 is performed to approach the best modeling parameters which reproduce
 the observations within their errors. By combining all
 non-asteroseismic constraints with asteroseismic measurements, we
 find that the observational constraints favour a model with \textbf{a mass of
 1.00$^{+ 0.01}_{- 0.02}$ $M_{\odot}$, an age t = 6.433 $\pm$ 0.04
 Gyr, a mixing-length parameter $\alpha$ = 1.75 $\pm$ 0.25 , an
 initial hydrogen abundance $X_{i}$ = 0.605$^{+ 0.01}_{- 0.005}$ and
 metal abundance $Z_{i}$ = 0.0275$^{+ 0.002}_{- 0.001}$}. \astrobjh{}
 is in post-main sequence phase of evolution. The modes of
 $l$ = 1 show up the characteristics of avoided crossings, which may be
 applied to test the internal structure of this type stars.
 Asteroseismic measurements can be used as a complementary
 constraint on the modeling parameters. \textbf{The models with mass
 1.00 - 1.10 $M_{\odot}$ can reproduce the observational constraints.
 Existing observed data of \astrobjh{} do not rule out these models.}

\end{abstract}


\end{frontmatter}

% main text
\section{Introduction}

Solar-like oscillations have been confirmed for several
main-sequence and subgiant stars, such as \astrobjca{}
\citep{bedd04}, \astrobjcb{} \citep{kjel05}, \astrobjp{}
\citep{egge04b}, \astrobjb{} \citep{carr05}, etc. The large and
small frequency separations of p-modes can provide a good estimate
of the mean density and age of the stars \citep{ulr86, ulr88}.  It
has been proven that asteroseismology is a powerful tool for
determining the fundamental parameters of the stars \citep{egge05,
egge06}.

\astrobjh{ (HR 6623, HD 161797, HIP 86974)} is a G5 IV subgiant
star, in the neighbourhood of the Sun. It is considered to be a
solar-type star with mass 1.14 $M_{\odot}$, effective temperature
$T_{e}$ = 5596 $\pm$ 80 K, and radius R = 1.77 $\pm$ 0.07
$R_{\odot}$ \citep{fuhr98}. Recently, \citet{bona08} detected
solar-like oscillations on $\mu$ Her and identified individual
frequencies in the range of 900 - 1600 $\mu$Hz. These seismological
data will provide a constraint on the fundamental parameters of
\astrobjh{}. In this work, we try to determine modeling parameters
of \astrobjh{} based on asteroseismic constraints using the Yale
Rotation Evolution Code (YREC7) in its non-rotating configuration.

The observational constraints available for \astrobjh{} are
summarized in Sect. 2, while the details of the evolutionary models
and the computational method are given in Sect. 3. The results are
presented in Sect. 4 and the conclusion is given in Sect. 5.

\section{Observational constraints}
\subsection{Effective temperature, luminosity and chemical composition}
The effective temperature of \astrobjh{} given by \citet{fuhr98} is
5596 K, however that given by \citet{ivan04} is only 5390 K.
Combining other data \citep{vald04, take05, soub08}, we adopt an
average effective temperature $T_{eff}$ = 5500 $\pm$ 90 K.

The luminosity of a star can be obtained through combining the
knowledge of the magnitude and distance. By combining the visual
magnitude V = 3.417 $\pm$ 0.014, the bolometric correction BC = -
0.15 $\pm$ 0.05 mag \citep{fuhr98}, the solar absolute magnitude
$M_{bol,\odot}$ = 4.746 \citep{leje98} and the newest Hipparcos
parallax $\Pi$ = 120.33 $\pm$ 0.16 mas \citep{leeu07}, we obtained a
luminosity for $\mu$ Her of L = 2.70 $\pm$ 0.16 $L_{\odot}$.

In the version of the Catalogue of [Fe/H] determinations given by
\citet{cayr01}, there are five metallicity values for $\mu$ Her.
Recent determinations give the values: 0.26 \citep{soub05} and 0.29
\citep{take05}. We adopt the average of these determinations, [Fe/H]
= 0.21 $\pm$ 0.07. This value is close to the value of 0.23 given by
\citep{fuhr98}. For Population I stars, the ratio of surface heavy
elements to hydrogen abundance is related to the Fe/H by $[Fe/H]
\simeq log(Z/X)_{s} - log(Z/X)_{\odot}$, where $(Z/X)_{\odot}$ is
the ratio of the solar mixture. The most recent ratio of the
heavy-element abundance to hydrogen abundance of the Sun,
$(Z/X)_{\odot}$, is 0.0171 \citep{aspl04}. There are, however, some
discrepancies between this new value and seismical results
\citep{yang07}. Thus in this work, we adopt the old value, 0.0245
\citep{grev93}. Consequently, the value of $(Z/X)_{s}$ for $\mu$ Her
is about 0.040 $\pm$ 0.006.

\subsection{Asteroseismic data}
solar-like oscillations for this star have been detected by
\citet{bona08} with the SARG echelle spectrograph. Twenty
oscillation frequencies have been identified by using modified and
standard extraction methods between 900 and 1600 $\mu$Hz. By means
of a least square best fit with the asymptotic relation of
frequencies for all the identified modes, \citet{bona08} gave that
the most likely value of the mean large and small frequency
separation is $\Delta\nu$ = 56.50 $\pm$ 0.07 $\mu$Hz and $\delta\nu$
= 5.03 $\pm$ 0.94 $\mu$Hz, respectively. Some of the observational
constraints for \astrobjh{} used in this work are given in
Table~\ref{tab1}.

\begin{table}
\begin{center}
\caption{Observational data for \astrobjh{} used in this work.
\label{tab1}}
\begin{tabular}{ll}
\hline
\hline

  $\Pi$ [mas]           &  120.33 $\pm$ 0.16    \\
  V [mag]                &  3.417 $\pm$ 0.014    \\
  $L/L_{\odot}$           &  2.70 $\pm$ 0.16     \\
  $T_{eff}$ [K]          &  5500 $\pm$ 90       \\
  $[Fe/H]_{s}$           &  0.21 $\pm$ 0.07       \\
  $\Delta\nu$ [$\mu$Hz]  &  56.50 $\pm$ 0.07      \\
  $\delta\nu$ [$\mu$Hz]  &  5.03 $\pm$ 0.94      \\
\hline

\end{tabular}
\end{center}
\end{table}

\section{Stellar models}
% \label{}
\subsection{Input physics}
A grid of stellar evolutionary models was computed with the YREC7 in
its non-rotating configuration \citep{guen92}. All models are
evolved from fully convective pre-main sequence (PMS) to a stage of
subgiant. The OPAL EOS2001 \citep{roge02}, OPAL opacity
\citep{igle96}, and the \citet{alex94} opacity for low temperature
were used. These opacity tables have the solar mixtures given by
\citet{grev93}. The models took into account diffusion of helium and
metals, using the prescription of \citet{thou94}. Energy transfer by
convection is treated according to the standard mixing-length
theory, and the boundaries of the convection zones are determined by
the Schwarzschild criterion. See \citet{dema07} in details for the
YREC.

\subsection{Computational method}
The position of a stellar model in the Hertzsprung-Russell (H-R)
diagram depends on five modeling parameters: the stellar mass $M$,
the mixing-length parameter $\alpha$, the age of the star $t$, the
initial hydrogen abundance $X_{i}$ and metallicity $Z_{i}$. For
$\mu$ Her we take the ratio of the heavy-element abundance to
hydrogen abundance on the stellar surface as an observable. Thus we
have three observables ($L, T_{eff}, (Z/X)_{s}$) and five unknowns
($M, \alpha, t, X_{i}, Z_{i}$).

In order to reproduce the observational constraints, we construct a
grid of models with various masses, initial element abundances and
mixing-length parameters. For each stellar model, low-degree p-mode
frequencies are computed using Guenther adiabatic pulsation code
\citep{guen94}. To find the set of modeling parameters ($M, \alpha,
Z_{i}, X_{i}, t$) that leads to the best agreement with the
observational constraints, following \cite{egge04a}, we perform a
$\chi^{2}$ minimization. The function $\chi^{2}$ is defined as
follows
\begin{equation}
 \chi^{2} = \chi^{2}_{clas} + \chi^{2}_{osci},
\label{equa1}
\end{equation}
where
\begin{equation}
 \chi^{2}_{clas} = (\frac{T^{eff}_{mod}-T^{eff}_{obs}}{\sigma(T^{eff}_{obs})})^{2}+
(\frac{\log(\frac{L}{L_{\odot}})_{mod}-\log(\frac{L}{L_{\odot}})_{obs}}
{\sigma(\log(\frac{L}{L_{\odot}})_{obs})})^{2}+
(\frac{[\frac{Fe}{H}]_{mod}-[\frac{Fe}{H}]_{obs}}{\sigma([\frac{Fe}{H}]_{obs})})^{2},
\label{equa2}
\end{equation}
and
\begin{equation}
 \chi^{2}_{osci} = \frac{1}{N}\sum^{N}_{i=1}
 (\frac{\nu^{theo}_{i} - \nu^{obs}_{i} - \langle
 D_{\nu}\rangle}{\sigma})^{2}.
\label{equa3}
\end{equation}
Here the $\sigma( )$'s are the errors on the corresponding
observations, $N$ is the number of observed frequencies, $\sigma$ =
1.8 $\mu$Hz is the resolution on the observed frequencies, and
$\langle D_{\nu}\rangle = \sum^{N}_{i=1}(\nu^{theo}_{i} -
\nu^{obs}_{i})/N$. The model which can minimize $\chi^{2}_{clas}$
and $\chi^{2}_{osci}$ at the same time will be considered to be the
best one.

\section{Results}
\begin{figure}
\begin{center}
\includegraphics[width=4cm,angle=-90]{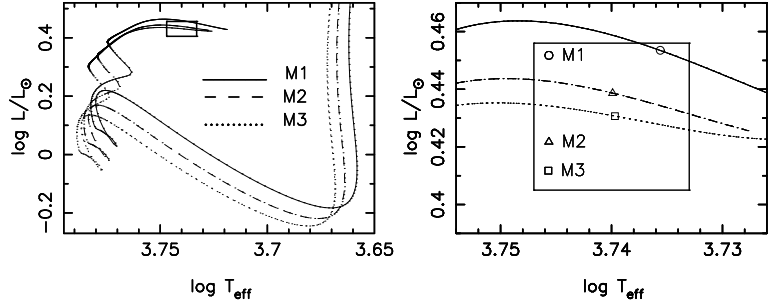}
\end{center}
\caption{Evolutionary tracks in the H-R diagram for models of
\astrobjh{}. Left panel: full tracks. Right panel: close-up of H-R
diagram in the vicinity of \astrobjh{}. The positions in the H-R
diagram of the models listed in Table \ref{tab2} are indicated. The
box shows the position of \astrobjh{}. } \label{fig1}
\end{figure}

\begin{figure}
\begin{center}
\includegraphics[width=6cm,angle=-90]{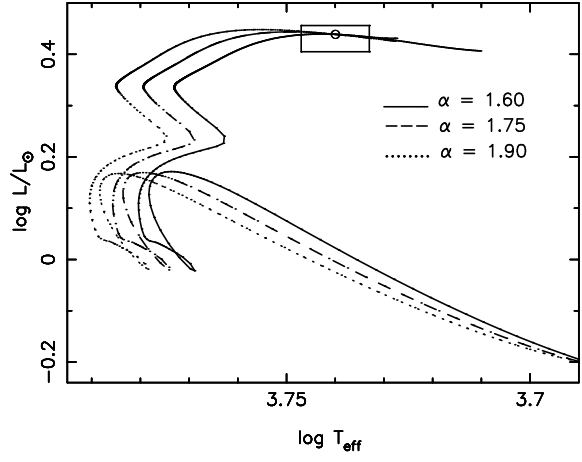}
\end{center}
\caption{Evolutionary tracks in the H-R diagram for models with
$Z_{i}$ = 0.0275, $X_{i}$ = 0.605, and M = 1.00 $M_{\odot}$, with
the lines differing in the value of the mixing-length parameter,
$\alpha$. The box shows the position of \astrobjh{}. Open circles
indicate the locations of the M2a, M2 and M2b models listed in Table
\ref{tab3}. } \label{fig2}
\end{figure}

% \label{}
\textbf{Firstly}, we computed a grid of evolutionary tracks for
models with masses, mixing length parameters, initial metal and
hydrogen mass fractions respectively in the ranges of  0.86
$M_{\odot}\leq M \leq$ 1.16 $M_{\odot}$, 1.5 $\leq \alpha \leq$
2.10, 0.025 $\leq Z_{i} \leq$ 0.033, 0.57 $\leq X_{i} \leq$ 0.70
\textbf{with a primary resolution $\delta M=0.02, \delta\alpha=0.2,
\delta Z=0.002$, and $\delta X=0.02$}. Computational results show
that the evolutionary tracks for models with $0.9 M_{\odot} < M <
1.15 M_{\odot}$ and $1.5 \leq \alpha \leq 2.10$ approximately span
the error box in [$T_{eff}, L, (Z/X)_{s}$]. This is because the fact
that a decrease of the mass can be compensated by a decrease in
hydrogen and metal abundances to get the same position in the H-R
diagram \citep{meng08}. Moreover, the evolutionary tracks imply that
$\mu$ Her is in the post-main sequence phase of evolution. In this
phase, the tracks are almost parallel to the $T_{eff}$-axis (see
Figs. \ref{fig1}, \ref{fig2}). Additionally, a variation of the
mixing length parameter mainly changes the radius, but has almost no
influence on the luminosity \citep{kipp90}. Therefore, with
increasing the mixing length parameter $\alpha$, the evolutionary
track moves almost horizontally to the left of H-R diagram and vice
versa. Thus the models with various mixing length parameters can
span the same position in the H-R diagram (see Fig. \ref{fig2}) at
different ages. Table \ref{tab3} gives the characteristics of three
models, M2a, M2 and M2b, with different mixing length parameters but
a same surface metallicity and position in the H-R diagram. These
models have different $\chi^{2}_{osci}$, i.e. different oscillation
frequencies, reflecting the differences in the internal structure of
the models; for example, the differences in the central helium-core
mass and density. This indicates that non-asteroseismic
observational constraints do not enable us to determine the
mixing-length parameter $\alpha$ for \astrobjh{}, an evolved
solar-type star, but asteroseismic observations could provide a
constraint on the mixing length parameter.

\begin{table}
\begin{center}
\caption{Models for \astrobjh{}. The superscripts or subscripts of
some modeling parameters show confidence limits. \label{tab2}}
\begin{tabular}{cc|ccc}
\hline \hline

              & $\log L/L_{\odot}$ &  \multicolumn{3}{c}{ 0.431 $\pm$ 0.025 }   \\
Observational & $T_{eff}$ [K]      &  \multicolumn{3}{c}{  5500 $\pm$ 90 }      \\
constraints   & $(Z/X)_{s}$        &  \multicolumn{3}{c}{ 0.040 $ \pm$ 0.006 }  \\
              & $\Delta\nu$        &  \multicolumn{3}{c}{ 56.50 $ \pm$ 0.07 }   \\
 \hline
           &              &  M1   & M2 & M3     \\
           & $M/M_{\odot}$& 1.10$_{- 0.01}$ & 1.00$^{+ 0.01}_{- 0.02}$ &0.93 $\pm$ 0.015\\
Modeling   & $\alpha$     & 1.75 $\pm$ 0.25&  1.75 $\pm$ 0.25& 1.75 $\pm$ 0.25\\
parameters & $Z_{i}$      & 0.029$^{+0.001}$  & 0.0275$^{+ 0.002}_{- 0.001}$&0.026 $\pm$ 0.002 \\
           & $X_{i}$      & 0.650$^{+0.005}$  & 0.605$^{+ 0.01}_{- 0.005}$&0.572 $\pm$ 0.01\\
\hline
              & $\log L/L_{\odot}$& 0.4550   & 0.4386 &0.4307  \\
              & $T_{eff}$ [K]     & 5454     & 5494   & 5492  \\
Model         & $Z_{s}$           &0.0272    & 0.0257 & 0.0245 \\
characteristics & $(Z/X)_{s}$     & 0.040    & 0.040  & 0.041  \\
              & $R/R_{\odot}$     &1.894     & 1.831  & 1.816  \\
              & age [Gyr]         & 6.556 $\pm$ 0.03 & 6.433 $\pm$ 0.04 & 6.303 $\pm$ 0.03 \\
              & $\chi^{2}_{osci}$ & 1.05     & 1.07   & 4.55  \\
              & $\chi^{2}_{clas}$ & 1.18     & 0.10   & 0.03  \\
              & $\chi^{2}$        & 2.23     & 1.17   & 4.58  \\
              & $\Delta\nu_{0}$   & 56.44    & 56.58  & 55.34  \\
               & $\delta\nu_{02}$ & 5.00     & 4.94   & 5.09  \\
 & $\langle D_{\nu}\rangle$ [$\mu$Hz]&28.10  & 31.20  & -0.53  \\
 & He-core mass [$M_{\odot}$]  &  0.100 &  0.101   & 0.105  \\
\hline
\end{tabular}
\end{center}
\end{table}

\begin{table}
\begin{center}
\caption{Characteristics of the models illustrated by open circles
in Fig.\ref{fig2}. \label{tab3}}
\begin{tabular}{cccc}
\hline \hline
                    & M2a  & M2  & M2b \\
   $M/M_{\odot}$    & 1.00  & 1.00   & 1.00 \\
   $Z_{i}$          &$\mathbf{0.0275}$  & $\mathbf{0.0275}$  & $\mathbf{0.0275}$ \\
   $X_{i}$          &0.605  &0.605   & 0.605\\
  $\alpha$          & 1.60  &  1.75  & 1.90  \\
\hline
     age [Gyr]      &  6.374 $\pm$ 0.04& 6.433 $\pm$ 0.04 & 6.471 $\pm$ 0.03  \\
  $\log L/L_{\odot}$& 0.4387 & 0.4386 & 0.4387 \\
  $T_{eff}$ [K]     & 5494  & 5494   & 5493  \\
  $(Z/X)_{s}$       & 0.040 &  0.040   & 0.041  \\
 $R/R_{\odot}$      &  1.832 &  1.831   & 1.832  \\
 $\chi^{2}_{clas}$  &  0.099 &  0.097   & 0.101  \\
  $\chi^{2}_{osci}$  &$\mathbf{3.50}$ &  1.07   & 3.37  \\
  $\Delta\nu_{0}$    & 56.65    & 56.58  & 56.46  \\
  $\log \rho_{c}$ [g/$cm^{3}$] &  3.636 & 3.800  & 3.942  \\
He-core mass [$M_{\odot}$]  &  0.091 &  0.101   & 0.109  \\
\hline

\end{tabular}
\end{center}
\end{table}

\begin{figure}
\begin{center}
\includegraphics[width=5cm,angle=-90]{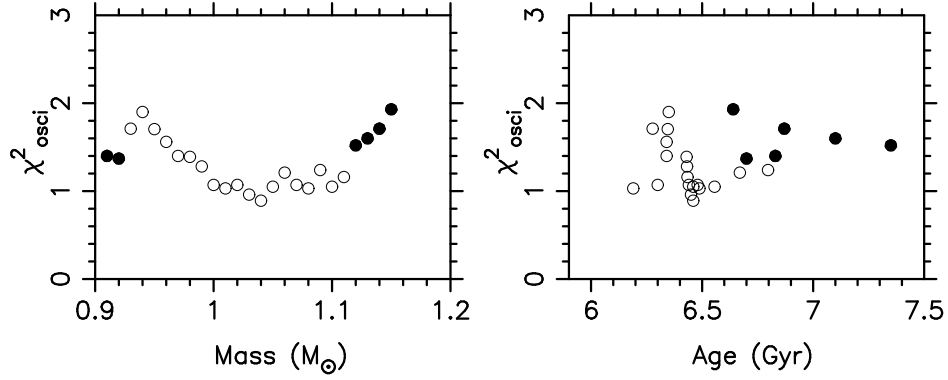}
\end{center}
\caption{Lowest values of $\chi^{2}_{osci}$ as a function of mass
and age. Open circles indicate models in accordance with the
classical observational constraints, while dots show models that do
not fall within the observational error box.} \label{fig3}
\end{figure}

\textbf{For the sets of modeling parameters that lead to agreements
with the observational constraints, we calculated the models with a
fine resolution $\delta M=0.01, \delta\alpha=0.05, \delta Z=0.0005$,
and $\delta X=0.005$ in the vicinity of the parameters. We obtained
many models that can almost minimize the $\chi^{2}_{osci}$ and fall
within the observational error box. Fig. \ref{fig3} shows the
$\chi^{2}_{osci}$ as a function of mass and age. From this figure we
can see that the models with mass 1 - 1.1 $M_{\odot}$ and age 6.2 -
6.7 Gyr can better reproduce the observed frequencies. We also
constructed the models with higher resolutions of modeling
parameters than the fine resolution. But results are not sensitive
to the resolutions.} Performing the $\chi^{2}$ minimization
described above, we found a solution: M = 1.00 $M_{\odot}$, $\alpha$
= 1.75, $Z_{i}$ = 0.0275, $X_{i}$ = 0.605, and t = 6.433 Gyr, marked
M2. Table \ref{tab2} lists characteristics of this model.
\textbf{The confidence limits of each modeling parameter correspond
to the maximum/minimum values it can reach when other parameters are
fixed, in order that the generated models fall within the
observational error box.} Corresponding evolutionary tracks are
shown in Figure \ref{fig1}. Although the value of $\chi^{2}_{clas}$
and $\chi^{2}_{osci}$ of the M2 model is not the lowest one
respectively, this model is almost able to minimize
$\chi^{2}_{clas}$ and $\chi^{2}_{osci}$ at the same time and has a
lowest $\chi^{2}$. \textbf{Fig. \ref{fig3} shows the model with a
mass of 1.04 $M_{\odot}$ has a lowest $\chi^{2}_{osci}$. But the
value of $\chi^{2}$ of this model is larger than that of M2.} The
mean large and small separation of the M2 model is 56.58 and 4.94
$\mu$Hz, respectively, which are in good agreement with those
observed. However, the mass of M2 model, 1.00 $M_{\odot}$, is less
than the values given by \cite{fuhr98} and \cite{take05}. They
showed that the mass of $\mu$ Her is about 1.14 $M_{\odot}$. Table
\ref{tab2} also gives a model M1 which has a mass as high as 1.10
$M_{\odot}$. This model is able to minimize $\chi^{2}_{osci}$ in the
age of 6.556 Gyr but has a higher value of $\chi^{2}_{clas}$. This
implies that increasing the hydrogen abundance and mass at the same
time for a fixed value of the mixing length parameter leads to an
increase in the $\chi_{clas}^{2}$ in order to attain the same value
of $\chi_{osci}^{2}$. This scenario was also found by \cite{egge06}.
The mean large and small separation of the M1 model is 56.44 and
5.00 $\mu$Hz, respectively, which are almost same as those of the M2
model.

\begin{figure}
\begin{center}
\includegraphics[width=5cm,angle=-90]{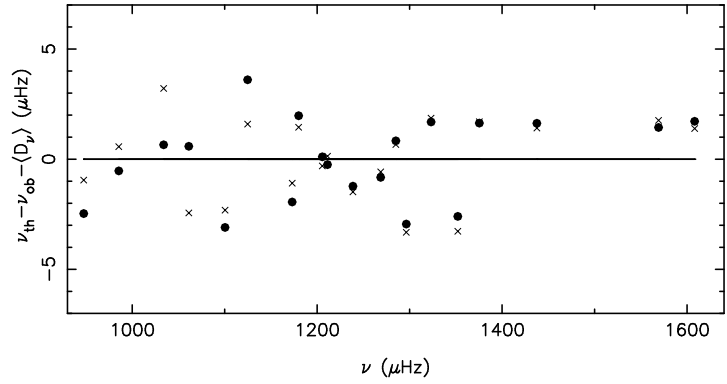}
\end{center}
\caption{Differences between computed and observed frequencies.
Crosses and filled circles correspond to the M1 and M2 model,
respectively. The systematic difference $\langle D_{\nu}\rangle$ has
been taken into account in this figure.} \label{fig4}
\end{figure}

\begin{figure}
\begin{center}
\includegraphics[width=4cm,angle=-90]{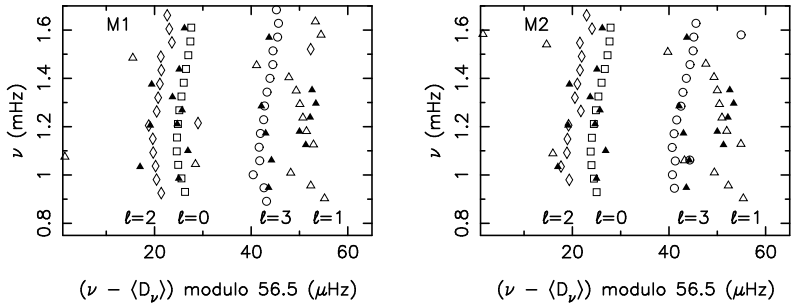}
\end{center}
\caption{Echelle diagrams for the M1 and M2 model, respectively.
Filled triangles refer to the observed frequencies \citep{bona08},
while open symbols correspond to the theoretical ones. Squares are
used for modes with $l$ = 0, triangles for $l$ = 1, diamonds for $l$
= 2 and circles for $l$ = 3.} \label{fig5}
\end{figure}

In Fig. \ref{fig4}, we plotted the differences between the observed
and theoretical frequencies for the M1 and M2 model. Moreover, the
theoretical and observed frequencies are compared by plotting
echelle diagrams of the M1 and M2 models in Fig. \ref{fig5}. The
systematic difference $\langle D_{\nu}\rangle$ between computed and
observed frequencies has been considered in these figures. The
systematic difference $\langle D_{\nu}\rangle$ is 28.1 and 31.2
$\mu$Hz for the M1 and M2 models, respectively. This systematic
difference also exists in other stellar models \citep{chri95,
egge04a, egge04b, egge06}. Fig. \ref{fig4} shows that the
differences between the observed and theoretical frequencies are
very similar for both models at high frequency, however they are
different at lower frequency. Compared to the M2 model, the M1 model
badly reproduces the observed frequencies of 1034.0 and 1061.2
$\mu$Hz, whereas the M2 model does not reproduce the observed
frequencies of 947.6 and 1124.8 $\mu$Hz well. These comparisons
between individual observed and theoretical asteroseismic
frequencies for the M1 and M2 models do not allow us to
differentiate which model is better. Additionally, both models have
the almost same values of the mean large and small separations
reflecting a similarity of the internal structure between the M1 and
M2 models. For example, both models have an almost same helium-core
mass.

\begin{figure}
\begin{center}
\includegraphics[width=6cm,angle=-90]{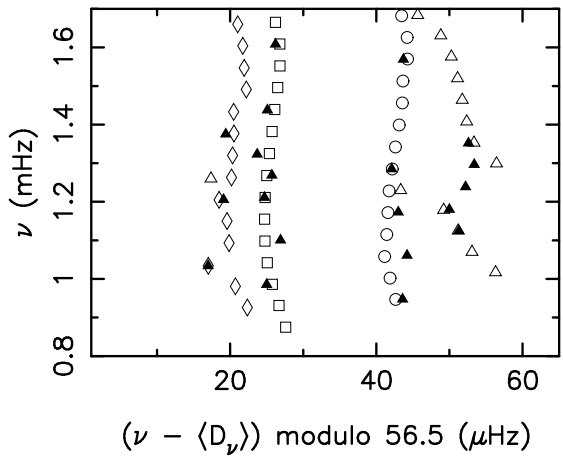}
\end{center}
\caption{Echelle diagrams for the M2b model. Filled triangles refer
to the observed frequencies \citep{bona08}, while open symbols
correspond to the theoretical ones. Squares are used for modes with
$l$ = 0, triangles for $l$ = 1, diamonds for $l$ = 2 and circles for
$l$ = 3. The systematic shift $\langle D_{\nu}\rangle$ is 28.4
$\mu$Hz for this model.} \label{fig6}
\end{figure}

\begin{figure}
\begin{center}
\includegraphics[width=9cm,angle=-90]{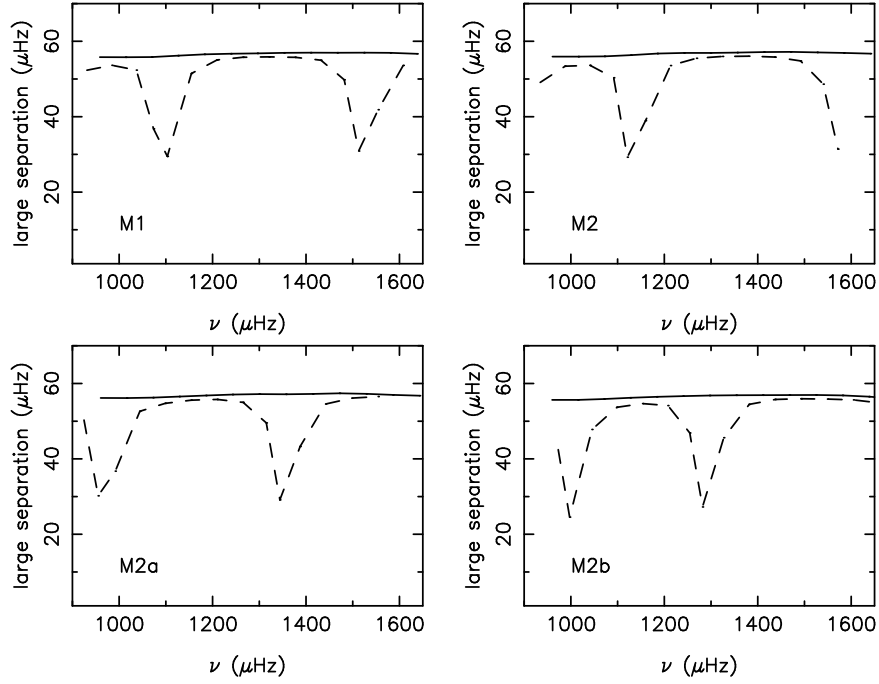}
\end{center}
\caption{ Large separations of theoretical frequencies of different
models. The solid line shows the large separations of the $l$ = 0
modes, while the dashed line indicates the large separations of the
$l$ = 1 modes.} \label{fig7}
\end{figure}

However, Fig. \ref{fig5} shows that both the M1 and M2 models badly
reproduce the observed frequencies of $l$ = 1. We searched the
models which can better reproduce the observed frequencies of $l$ =
1. A model labeled M2b, which was shown in Table \ref{tab3}, was
found. However, the value of $\chi^{2}_{osci}$ for the M2b model is
larger than \textbf{those} of the M1 and M2 models, which is mainly
due to differences in the modes of $l$ = 1 of these models. If the
contribution to $\chi^{2}_{osci}$ from the $l$ = 1 mode near 1238
$\mu$Hz is neglected, the value of $\chi^{2}_{osci}$ for the M2b
model will decrease from 3.37 to 0.72. Echelle diagram for the M2b
model is shown in Fig. \ref{fig6}. Except for the $l$ = 1 mode near
1238 $\mu$Hz, the M2b model reproduces the observed frequencies
well. Moreover Fig. \ref{fig6} shows that the observed and
theoretical frequencies with $l$ = 1 have a similar behavior:
\textbf{a zigzag echelle diagram}. The modes of $l$ = 1 show up a
deviation from their expected asymptotic values. Fig. \ref{fig7}
also shows that the large separations for modes of $l$ = 1 deviate
from their expected asymptotic behaviour. This may be because that
the formation of the central helium core leads to an increase in the
frequencies of the g-modes. `When the frequency of a g-mode
approaches that of a p-mode, the two modes undergo an avoided
crossing, where they exchange physical nature' \citep{aize77,
chri95}. Figs. \ref{fig6} and \ref{fig7} show that the $l$ = 1 modes
of the M2b model between 1200 and 1300 $\mu$Hz may undergo the
avoided crossing. Noting the helium-core mass of the M1 and M2
models are almost same, but that of the M2 and M2b models are
different, the behaviour of the $l$ = 1 modes might be very
sensitive to the internal structure of the stars. It may be applied
to extract the information of the helium core.

\begin{figure}
\begin{center}
\includegraphics[width=4cm,angle=-90]{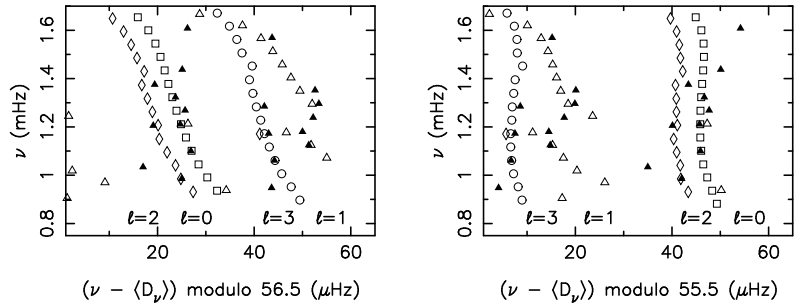}
\end{center}
\caption{Same as Fig. \ref{fig5}, but for the M3 model. Left panel
shows frequencies modulo the value of observed large separation,
while right panel presents frequencies modulo the value of
theoretical large separation.} \label{fig8}
\end{figure}

We also considered a $\chi^{2}_{osci}$ which does not contain
$\langle D_{\nu}\rangle$. For models falling within the observed
constraints on effective temperature, luminosity and surface
metallicity, we computed this new $\chi^{2}_{osci}$. A solution with
M = 0.93 $M_{\odot}$, $\alpha$ = 1.75, $Z_{i}$ = 0.026, $X_{i}$ =
0.572, and t = 6.303 Gyr was found. The evolutionary track and the
position of this model (marked M3) in the H-R diagram are also shown
in Fig. \ref{fig1}. The characteristics of the M3 model are given in
Table \ref{tab2}. The systematic difference $\langle D_{\nu}\rangle$
between computed and observed frequencies is -0.53 $\mu$Hz for this
model. However the value of the new $\chi^{2}_{osci}$ is 4.55, which
is much larger than that of the M1 and M2 models. Moreover the
echelle diagram of the M3 model is plotted in Fig. \ref{fig8}. This
echelle diagram shows that the M3 model can reproduce the
frequencies in the range of 1034 $< \nu <$ 1437 $\mu$Hz but badly
reproduces the modes at the lower and higher frequencies. Moreover
its mean larger separation is only 55.49 $\mu$Hz. Thus this model is
in disagreement with the asteroseismic constraints.

\section{Discussion and Conclusion}
% \label{}

The mass of 1.00 $M_{\odot}$ for the M2 and M2b models is less than
the value in the previous literatures \citep{fuhr98, take05}, which
may result from that the results of \cite{fuhr98} and \cite{take05}
were obtained without the asteroseismic constraints. The initial
hydrogen mass fraction is 0.605 and 0.650 for the M2 and M1 model,
respectively. Although there is a difference of 0.0015 in Z between
the M1 and M2 model, the increase of the hydrogen or decrease of
helium is mainly compensated by an increase of the mass in order to
reach the same location in the H-R diagram: a helium-mass degeneracy
\citep{lebr93}. The M1 and M2 models have the same value of
$\chi_{osci}^{2}$ and the almost same asteroseismic features. This
implies that the helium-mass degeneracy is difficult to be removed,
even if the asteroseismic constraints are taken into account.

Comparing with the observed frequencies, the theoretical frequencies
have a systematic shift. \textbf{The mode of $l$ = 1 near 1238
$\mu$Hz deviate from its expected asymptotic value in Fig.
\ref{fig6}. It may undergo an avoided crossing and be a mixed-mode.
The mixed-mode could have higher inertia than other p-modes. The
shift of this mode may be different from that of other p-modes. Thus
we divide the $\chi_{osci}^{2}$ into two parts: one
$\chi_{oscm}^{2}$ for mixed-modes and one $\chi_{oscp}^{2}$ for all
modes except for mixed-modes. Assuming only the $l=1$ mode near 1238
$\mu$Hz is a mixed-mode, we obtained the value of this new
$\chi_{osci}^{2}$ is 1.07, 1.10, 3.61 and 0.72 for M1, M2, M2a and
M2b, respectively. Under this assumption, M2b has a smallest
$\chi_{osci}^{2}$. }

We confirmed the results that an analysis of the H-R diagram does
not allow us to determine the mixing length parameter for an evolved
solar-type star \citep{fern03} but the observed oscillation
frequencies could provide a constraint on this parameter. The
internal structure of the evolved solar-type stars is sensitive to
the mixing length parameter at given ($T_{eff}, L$).

In this work we constructed the models for the \astrobjh{} using the
Yale stellar evolution code. By combining the non-asteroseismic
constraints with the asteroseismic observations, we find that a
model for $\mu$ Her can reproduce the all non-asteroseismic and
asteroseismic constraints well: \textbf{the model with a mass of
1.00$^{+ 0.01}_{- 0.02}$ $M_{\odot}$, an age t = 6.433 $\pm$ 0.04
Gyr, a mixing-length parameter $\alpha$ = 1.75 $\pm$ 0.25 , an
initial hydrogen abundance $X_{i}$ = 0.605$^{+ 0.01}_{- 0.005}$ and
metal abundance $Z_{i}$ = 0.0275$^{+ 0.002}_{- 0.001}$. However, the
models with mass 1 - 1.1 $M_{\odot}$ and age $6.2 - 6.7$ Gyr also
can reproduce the non-asteroseismic and asteroseismic constraints.
Existing observational constraints do not rule out those models.}

The modes of $l$ = 1 show up the characteristic avoided crossings,
which may be applied to test the internal structure of an evolved
solar-type star. The asteroseismic observations put important
constraints on the models for \astrobjh{}, but they are not enough
to really test the differences in the models. More accurate
oscillation frequencies, especially the modes of $l$ = 1, are need
to investigate the internal structure of this type star.

% Bibliographic references with the natbib package:
% Parenthetical: \citep{Bai92} produces (Bailyn 1992).
% Textual: \citet{Bai95} produces Bailyn et al. (1995).
% An affix and part of a reference:
%   \citep[e.g.][Ch. 2]{Bar76}
%   produces (e.g. Barnes et al. 1976, Ch. 2).

\end{document}